# The ground state of a quantum critical system

Wouter Montfrooij[1], Jagat Lamsal[1], Meigan Aronson[2], Marcus Bennett[2], Anne de Visser[3], Huang Ying Kai[3], Nguyen Thanh Huy[3], Mohana Yethiraj[4], Mark Lumsden[4], Yiming Qiu[5]

[1] Department of Physics and Astronomy, University of Missouri, Columbia MO
[2] Department of Physics, University of Michigan, Ann Arbor MI
[3] Van der Waals-Zeeman Institute, University of Amsterdam, Amsterdam, the Netherlands
[4] Oak Ridge National Laboratory, Oak Ridge TN
[5] National Institute of Standards and Technology, Gaithersburg MD and University of Maryland, College Park, MD.

Abstract: **The competition between the tendency of magnetic moments to order at low temperatures, and the tendency of conduction electrons to shield these moments, can result in a phase transition that takes place at zero Kelvin, the quantum critical point (QCP). So far, the ground state of these types of systems remains unresolved. Here we show that the ground state of a sample representative of a class of QCP-systems is determined by the interactions between the conduction electrons, resulting in a state with incommensurate intermediate-range order. However, long-range order is thwarted by quantum fluctuations that locally destroy magnetic moments.**

The quantum critical point refers to the order-disorder phase transition that takes place at zero Kelvin as a function of tuning parameter, such as using pressure or applying magnetic fields (1,2). In a metal, this tuning changes the interaction between the atomic magnetic moments and the conduction electrons, driving a magnetically ordered state into a quantum-disordered state (3). At the QCP the interaction strength is critical: on the one hand there is the tendency of the magnetic moments to order, on the other hand the conduction electrons are now so strongly coupled to these moments that they become almost localized, thereby shielding the moments from each other. These semi-localized electrons are believed to be the root of the heavy-fermion state (4). Some of the QCP-systems exhibit E/T-scaling (5-8), the property that the response depends only on the ratio of probing energy E and temperature T. This E/T-scaling is one of the most puzzling phenomena in these strongly correlated electron systems as this scaling should not be allowed on general theoretical grounds (1) unless under special circumstances (9) that do not appear to have been met in some of the systems (6,8) where scaling is present.

Often, before reaching the QCP, the system finds a state of lower energy, such as a superconducting state made up of obese Cooper pairs (10,11), or even states where ferromagnetism and superconductivity coexist (12,13). Understanding the state of matter near the QCP where textbook (14) Fermi-liquid theory for metals fails might even be the starting point for explaining the phenomenon of high-temperature superconductivity (15). In order to understand the excited states of QCP-systems that determine properties such as superconductivity, we must first determine the ground state. In here, we show that the ground state of Ce(Ru$_{1-x}$Fe$_x$)$_2$Ge$_2$ when prepared to be at the QCP [x=0.76 (8,16)] is given by a spin-density wave (SDW); this SDW (17,18) would have spanned the entire sample had it not been for the random quantum fluctuations. Our findings show, for the first time, how local random effects shape the response, and how a SDW can be reconciled with E/T-scaling. The quantum fluctuations we observe in this system are relevant to all QCP-systems.

We performed neutron scattering experiments to determine the ground state of Ce(Ru$_{1-x}$Fe$_x$)$_2$Ge$_2$: a neutron interacts both with the atomic nucleus and with the unpaired electrons. The scattered intensity yields the positions of atoms and the orientation of their magnetic moments. Thus, we can determine over what length scales the moments are correlated along various crystallographic directions. Also, since a neutron moves slowly compared to an electron, the signal yields the size of the volume the electrons occupy during the scattering process, distinguishing d-electrons from f-electrons.

We used an 11 g Ce(Ru$_{1-x}$Fe$_x$)$_2$Ge$_2$ single crystal prepared to be at the QCP. Ce(Ru$_{1-x}$Fe$_x$)$_2$Ge$_2$ crystallizes in the ThCr$_2$Si$_2$ structure (16), with the magnetic cerium atoms forming a body centered tetragonal lattice (a= 4.1 Å, c= 10.4 Å). Results for sample homogeneity, resistivity, specific heat and susceptibility are presented in the supporting online material (SOM). The neutron scattering experiments were done using the HB3 triple-axis spectrometer at Oak Ridge National Laboratory, and the DCS time-of-flight spectrometer at the National Institute of Standards and Technology. Powder patterns were collected at the Missouri Research Reactor. The sample was mounted in the *hhl*-scattering plane allowing us to determine correlation lengths along the major crystallographic directions. The lowest temperature for the DCS (HB3) experiment was 0.4 K (1.56 K), using an incident neutron energy of 3.55 meV (14.7 meV), giving an energy resolution (full width at half maximum-FWHM) of 0.10 meV (1.25 meV).

The experiments show that Ce(Ru$_{1-x}$Fe$_x$)$_2$Ge$_2$ is a system with local moments associated with unpaired cerium f-electrons (19) that persist down to the lowest temperatures (Fig. 1). These moments are aligned along the c-axis (SOM). On lowering T, magnetic scattering increases (Fig. 1) around the incommensurate positions (n,n,2m±0.45), indicating that the cerium moments are on the verge of forming a spin-density wave with propagation wave vector 2**k**$_F$ = (0,0,0.45) (with **k**$_F$ the Fermi wave vector). In agreement with polycrystalline data (8,19), we find that at low T the moment sizes and directions (up/down) have become correlated with each other over a sizeable volume (~2x10$^4$ Å$^3$ at T= 2 K).

It is straightforward to isolate the magnetic scattering associated with the approach to an ordered state (SOM). In Fig. 2 we plot as a function of reciprocal lattice units (rlu) the net intensity measured at T= 2 K along 4 directions at the ordering wave vector (1,1,0.45). Here (and in the following) we have subtracted the signal at T= 56 K in order to eliminate incoherent nuclear scattering and the weak scattering by the cryostat and sample mount. While this procedure slightly overestimates the background (SOM), it does not complicate the data interpretation. Fig. 2 shows that there is a sharp, ordered component on top of the broader distribution associated with intermediate-range order. We attribute this sharp component to a Fe-poor phase at one end of the sample (SOM). Since the overall contribution (the area under the curve) of this phase is small, it does not affect our interpretation.

The HB3 spectra are described (SOM) by Lorentzian lineshapes in reciprocal space (Fig. 2), corresponding to correlations that decay in real space as $\sim\exp(-r/\xi)/r$. Remarkably, the correlation length $\xi$ ($\xi= 4\pi$/FWHM, with FWHM the full width at half maximum of the Fig. 2 spectra) in the [00$\eta$]-direction is identical to that measured in the [$\eta\eta$0]-direction, *provided* it is expressed in lattice units (lu). At 2 K where we took the most accurate data, the width of the spectra is 0.166 ± 0.003 rlu in the [00$\eta$]-direction and 0.166 ± 0.004 rlu in the [$\eta\eta$0]-direction. This identity holds true for all measured temperatures (Fig. 3), implying that in real space the moment correlations persist over almost 4 times larger distances (measured in Å) in the [00$\eta$]-direction than in the [$\eta\eta$0]-direction. Yet, the distance between Ce ions is actually larger along the [00$\eta$]-direction than it is along the [$\eta\eta$0]-direction. Thus, the spatial extent of the correlations cannot be accounted for by the strength of the intermoment coupling, irrespective of the origin of this coupling. If the coupling is conduction electron mediated (the most likely case), then the strength of this coupling (1) is expected to decrease as $1/r^3$ implying a much larger correlation length (in lu) along [$\eta\eta$0] than along [00$\eta$]. If instead the coupling originates from an indirect exchange mechanism, then the correlation length (in lu) along [$\eta\eta$0] should be (at least) double that of the correlation length along [00$\eta$]. Clearly, this is not compatible with our findings.

The DCS experiments show that the correlation length remains finite down to the lowest temperatures. While the DCS experiments were setup to measure the decay time of magnetic fluctuations, we can still determine a reasonable estimate for the low temperature correlation length. Because the DCS detectors are fixed in place, the signal for elastically scattered neutrons corresponds to a curved path in *hhl*-space. Along this path, the variation of *l* is much larger than the variation of *h*. We can estimate a correlation length from this if we assume that the [$\eta\eta$0] and [00$\eta$]-directions are equivalent. We plot the scattered intensity as a function of d= $\sqrt{[(h-1)^2+(l-0.45)^2]}$ in Fig. 3b. While the data collapse onto a single curve, note that d is largely determined by the variation in *l* so this collapse is not a justification of our assumption. As expected upon approaching a phase transition, a continued increase in scattering is observed around the ordering wave vector for T<2 K: in the units of Fig. 3b, we find for d= 0 a net signal strength of 6.4 ± 0.5, 4.2 ± 0.6, 2.9 ± 0.7 and 2.2 ± 0.5, at T= 0.4, 0.9, 1.2 and 1.8 K, respectively. The temperature dependent correlation lengths are shown in Fig. 3c, indicating that they remain finite down to the lowest temperature. The fact that the correlation length stays finite is likely caused by most of our sample being in the paramagnetic phase (SOM), however, it does show that we are in the part of the phase diagram where quantum fluctuations dominate thermal fluctuations. At T=0.4 K, the width *in energy* of the fluctuations associated with adopting a SDW ground state in the DCS experiment has become spectrometer resolution limited, implying that fluctuations linked to ordering take at least 80 ps to relax. The last piece of information that is needed to characterize the ground state is that the correlation length (in lu) along [ηηη] is 1.5 times the correlation length measured along [00η] (Fig. 2).

The findings listed above identify the fluctuations that drive quantum critical $Ce(Ru_{1-x}Fe_x)_2Ge_2$ away from ordering as quantum fluctuations associated with the zero-point motion of the Ce ions and/or concentrational fluctuations. We illustrate our reasoning in Fig. 4 where we have drawn volumes in real space that have the same symmetry as the Wigner-Seitz (WS) unit cell of the body centered *cube*. If we assume that there is a certain amount of correlation loss (e.g., 10%, the actual number is irrelevant) between the center of a WS-volume and the center of a neighboring volume, then one would expect an identical correlation loss along the [00η] and [ηη0]-direction, provided distances are measured in units corresponding to the spacing between planes. Also, along the [ηηη]-direction we would expect a correlation length that is 1.5 times larger (when measured in plane spacings) than the correlation length in the [00η]-direction. Thus, our observed correlation lengths reproduce the symmetry of a *cube*.

However, the observed correlation lengths are not in agreement with the tetragonal symmetry of our sample. The explanation for this is that the loss of correlation must occur at the cerium sites, irrespective of how far apart they are. The only mechanism that can explain this is a non-constant (in time or in space) overlap of the cerium f-orbitals with the conduction electron states. A constant interaction strength (3) would either be strong enough for the system to order (or not), but it cannot account for the observed correlation length being independent of the distance between the Ce ions. Bearing in mind that the system is tuned to have critical interaction strength (overlap), even a small change in overlap will have a noticeable effect (3). A spatially varying overlap can originate from local disorder: either an Fe or a larger Ru ion will occupy a specific lattice site. A temporally varying overlap originates from the zero-point motion of the cerium ions in their potential wells. The amplitude of this motion is typically of the order $(m/M)^{1/4}d$ (20), with m the electron mass and M the ion mass and d the lattice spacing. Independent of its actual amplitude [~0.1 Å at 16 K (SOM)], in the vicinity of the QCP this motion will be large enough to upset the critical balance between local moment shielding and moment surviving. Likely both effects conspire to give a non-constant interaction (SOM). Close to the QCP, zero-point motion will always have an effect, while lattice disorder (substitution, vacancies) will vary from sample to sample. We refer to both temporal and spatial fluctuations as quantum fluctuations.

The ordering characteristics resemble that of a percolation problem (21). Whether long-range order is achieved depends on whether there is a path linking moments on either side of the sample via moments that have permanently survived. Since the latter is determined at random, we cannot expect any directional preference, and a cubic symmetry of the correlation volume is the only possibility. Given this, and given the very slow time scale of the fluctuations representing SDW-order, we conclude that in our sample quantum fluctuations have prevented the system from adopting a long-range ordered SDW-ground state. Also, these findings give insight in how E/T-scaling is possible in a system whose effective (1) dimensionality (space+time) $D+z = 3+2$ is apparently above its critical dimension $D_c=4$. At the percolation threshold the lattice consists of long chains of interacting spins (21), the longest of which span the entire sample. This effectively reduces the spatial dimensionality of the system from D=3 to D=$1^+$. In addition, the random fluctuations in the degree of shielding of the moments could well be mimicked by applying a random magnetic field to a classical system. It is known (22) that such random fields increase the upper critical dimension. Individually, or combined, these two effects reduce the effective dimensionality of the system to below $D_c$, setting the stage for E/T-scaling.

In conclusion, we have identified the ground state of $Ce(Ru_{1-x}Fe_x)_2Ge_2$, a member of a class of intensely studied 122-quantum critical systems (1). On lowering the temperature, the conduction electrons couple to the local cerium moments, and thereby couple the local moments to each other. The local moments become increasingly more shielded by the conduction electrons due to the strong interaction between them. However, this shielding is not complete, and (some) local moments survive down to zero Kelvin. These local moments should not be viewed as unchanging; the quantum fluctuations that change the distances between a Ce ion and its neighbors change the interaction strength between the local moments and the conduction electrons, thereby altering the degree of shielding. The residual interactions between the almost localized conduction electrons induce the formation of a spin-density wave, determined by the morphology of the Fermi-surface. Had it not been for quantum fluctuations, this SDW would have spanned the entire crystal. In fact, even the smallest reduction in coupling between the local moments and the conduction electrons, which can be achieved by further expanding the crystal by substituting Ru atoms on the Fe sites, would allow for enough of the cerium moments to survive at all times for long range order to be established.

Acknowledgments: we thank Chris Redmon and Ross Erwin for running the cryogenic equipment and T. Gortenmulder for the EPMA work. The work was partly supported by the NSF [DMR-0405961 and DMR-0454672] and by the Dutch foundation FOM. ORNL is managed by UT-Battelle, LLC, for the US Department of Energy under contract No. DE-AC00OR22725.

Fig. 1. The onset of incommensurate ordering at low temperatures measured on HB3. There is a clear difference in scattering between 2 K and 56 K along [11η] (black solid symbols), whereas no difference is observed along [00η] (red, open symbols), showing that the moments are aligned along the c-axis. The yellow area indicates the magnetic scattering. The gaps in the data points are caused by the (1,1,2n) nuclear Bragg peaks and by the powder peaks due to the Al-sample holder. The solid blue line is the expected magnetic intensity based on the cross-section for neutrons and the Ce f-electron form factor (23), the dotted line is calculated for Fe d-electrons (SOM). The arrows are at $(1,1,2n \pm 0.45)$

Fig. 2. Net elastic neutron scattering intensity near the ordering wave vector (1,1,0.45) at T=2 K in four reciprocal space directions. The solid curves are fits to a Lorentzian line shape. The short horizontal lines at the origin of each figure denote the resolution FWHM determined at (1,1,0).

Fig. 3. (a) The net magnetic scattering measured on HB3 along [00η] (black symbols) and [ηη0] (red symbols), centered on (1,1,0.45). At all T, the FWHM in reciprocal space is identical along both directions. For comparison, a constant background has been added to the [ηη0] data, and they have been multiplied by 1.3 to allow for a direct comparison between the two directions despite the sharp feature observed in [ηη0] data (due to a Fe-deficient section).

(b) The uncorrected DCS data at 0.4 K taken for elastically scattered neutrons (E=0 ± 0.02 meV). The background is determined by the incoherent nuclear scattering. The solid line is a Lorentzian lineshape with FWHM of 0.14 rlu. The sharp feature at d=0 is ascribed to the small Fe-deficient part.

(c) The temperature dependence of the FWHM of the neutron scattering spectra. The HB3 data measured along [ηη0] ([00η]) are given by the open (solid) circles, the DCS data by the blue circles. The red symbols stand for initial HB3 experiments, black circles for follow up measurements. The line guiding the eye is ~ $\coth(\Delta/T)$, with $\Delta = 2.4$ K.

Fig. 4. The observed symmetry of the magnetic correlations in real space. Measuring distances in units of the corresponding lattice spacings ($d_{100}$, $d_{001}$, $d_{110}$ and $d_{111}$), steps of sizes $d_{001}$, $d_{100}$ and $d_{110}$ result in the same amount of correlation loss, while body-diagonally-sized steps (the length of the diagonal is $3d_{111}$) correspond to double the loss in correlation, or only 2/3 per $d_{111}$ spacing.

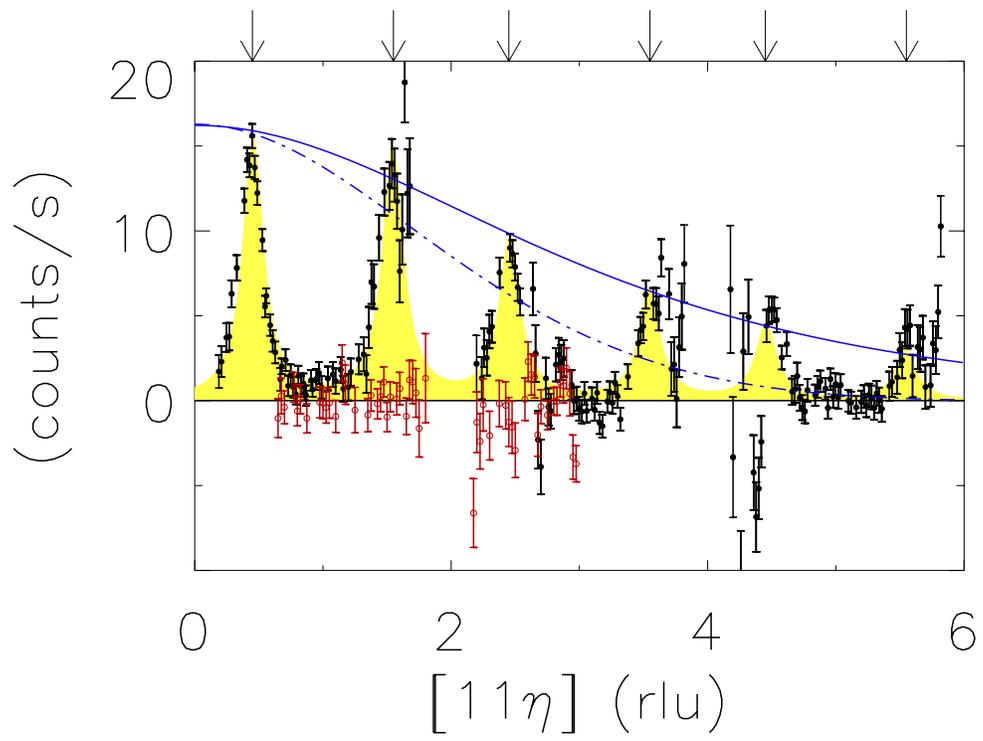

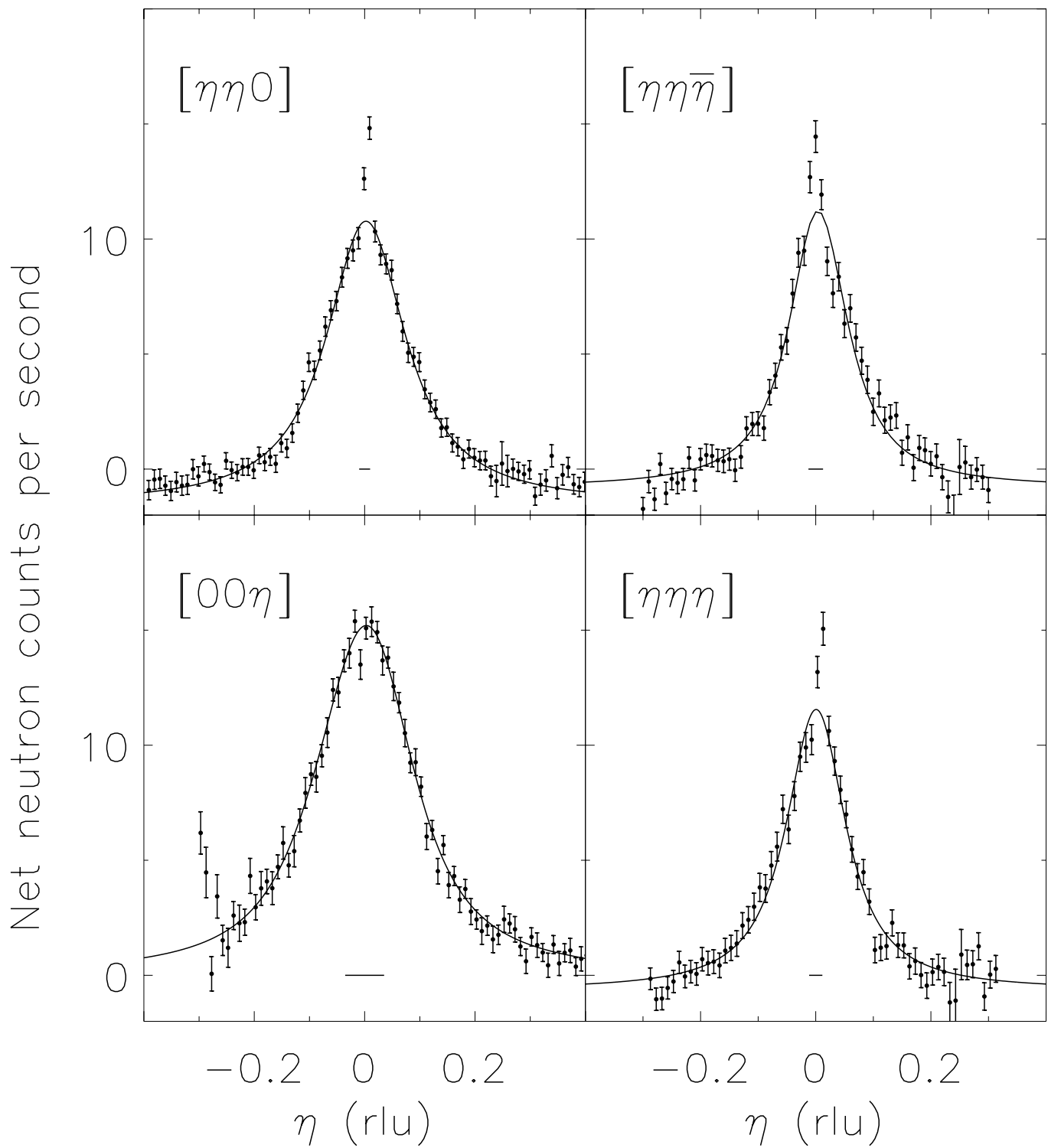

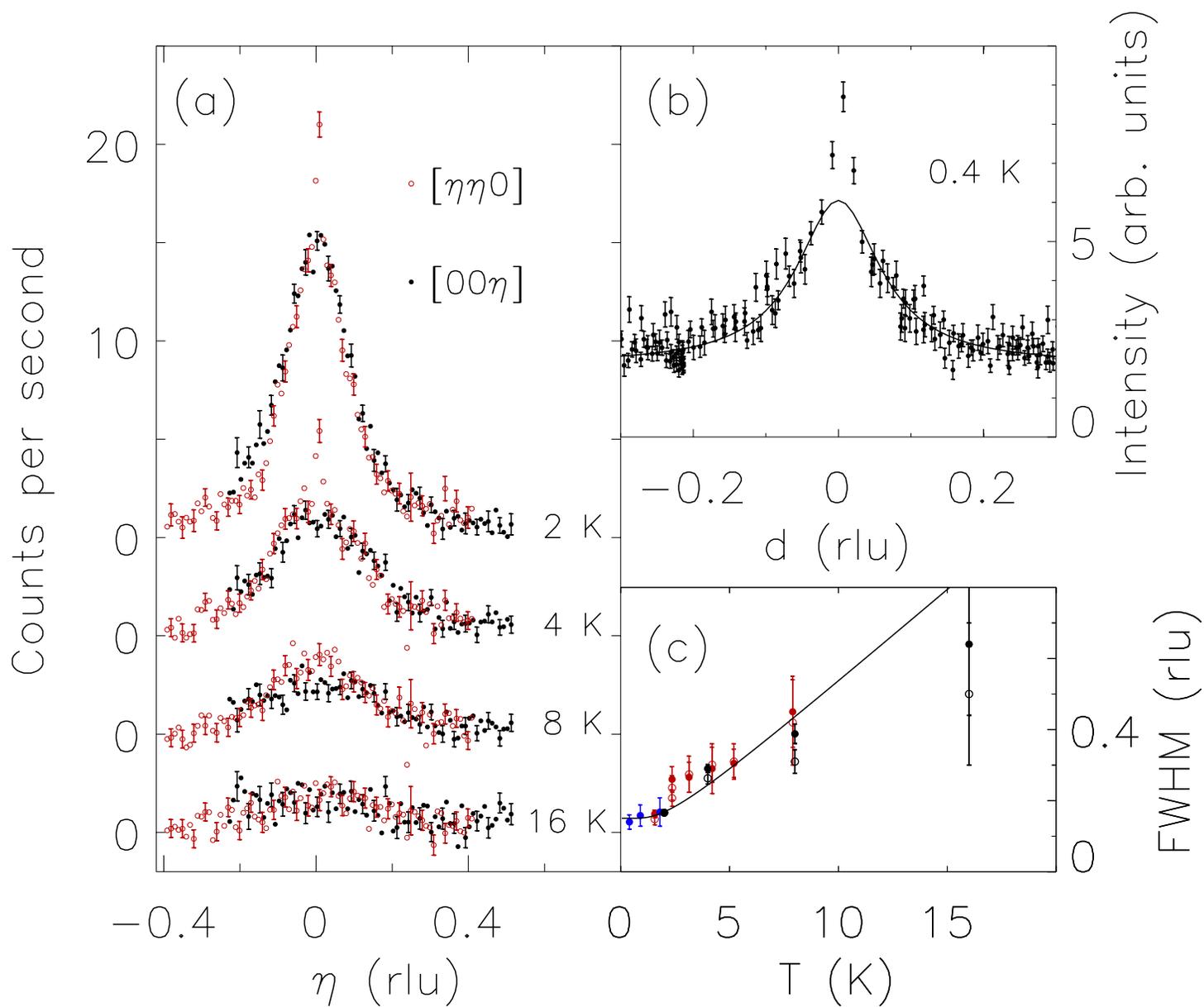

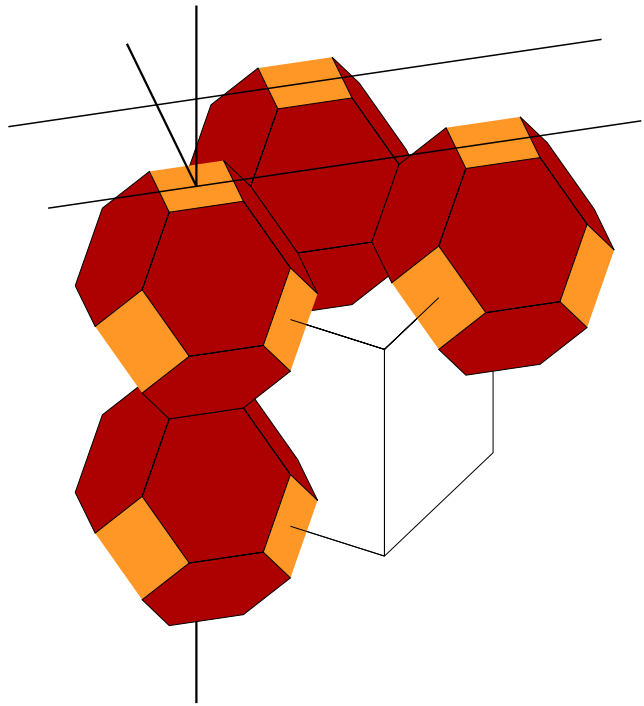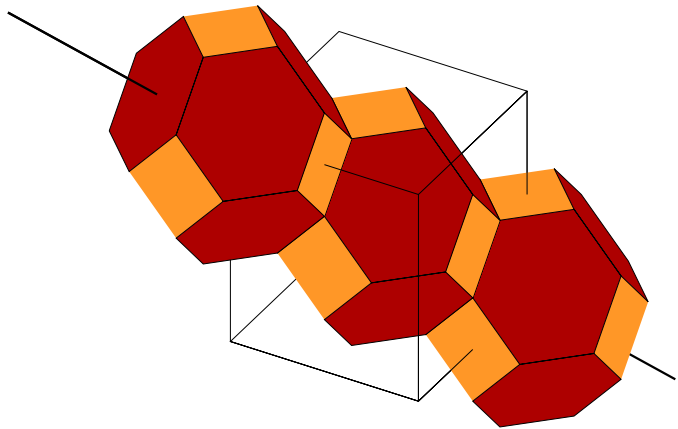

Supporting online material (SOM)

Sample preparation, quality and bulk measurements

The Ce(Ru$_{1-x}$Fe$_x$)$_2$Ge$_2$ single crystal ($x$= 0.76) was grown using a floating zone furnace yielding a 5 cm long cylindrical crystal of ~7 mm diameter with the [110] direction 5 degrees from the cylindrical axis. [Starting materials: Ce (3N5), Ru (3N5), Ge (5N), Fe (4N8)]. Higher Ru-concentrations yield a sample that orders at the lowest temperatures. The bottom 1 cm of the crystal was found to be slightly Fe-deficient, and was masked in the initial HB3 neutron scattering experiments, and cut off for the subsequent HB3 and DCS experiments. The remainder 11 g crystal did not show Bragg peaks indicative of polycrystallinity (in neutron scattering experiments or in Laue backscattering), however, about 3% of the crystal still turned out to be Fe-deficient (based on the area of the resolution limited component seen in neutron scattering experiments at T=2 K).

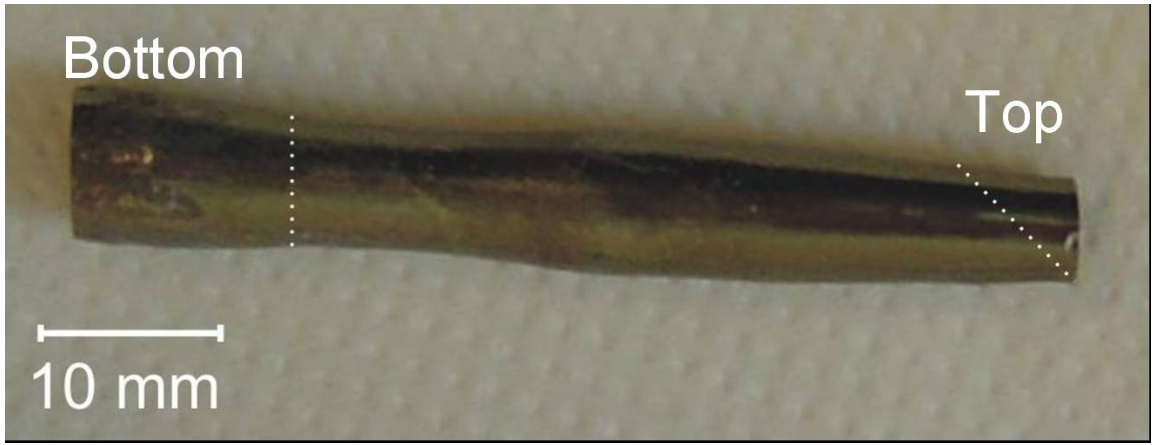

Fig. S1: Photograph of the single crystal as used in the initial HB3 experiment. For the initial HB3 experiment, the bottom part below the dotted line was masked with Gd paint, for subsequent DCS and HB3 experiments, the crystal was cut along the dotted lines.

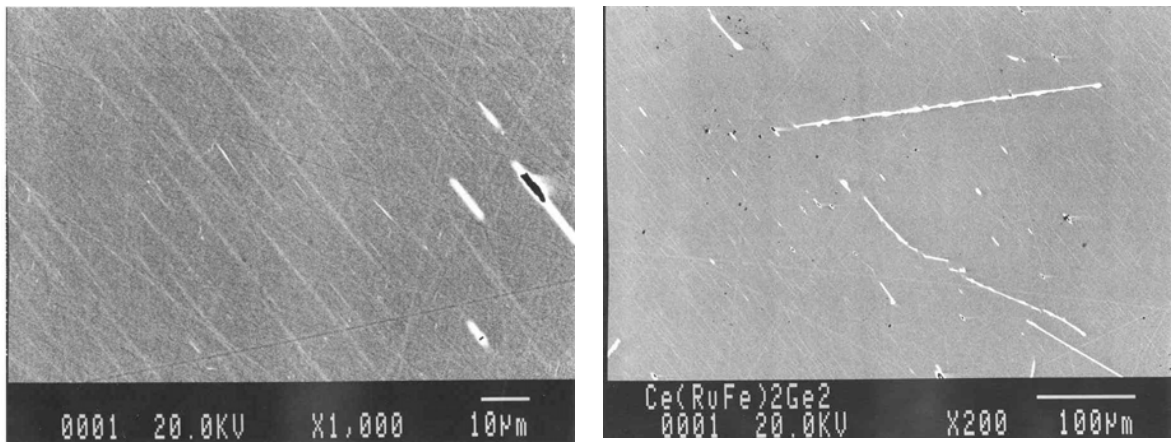

Fig. S2: Electron probe pictures of the bottom end (dotted line in Fig. S1) of the sample.

The cut-off pieces were used for the characterization, specific heat, resistivity and susceptibility measurements. The sample composition was determined using the electron probe microanalyzer (EPMA) JEOL JXA-8621. We found the compositions to be mainly stoichiometric with a few percentages of Ge-rich secondary phases. The pictures (Fig. S2) show the secondary phases as white stripes. Both ends of the crystal yielded very similar pictures. The sample composition (normalized to 5 atoms per formula unit) determined at either end of the crystal is $Ce_{0.988}(Ru_{0.233}Fe_{0.777})_2Ge_{1.993}$ for the top part of the sample and $Ce_{0.996}(Ru_{0.261}Fe_{0.753})_2Ge_{1.978}$ for the bottom part. Based on the phase diagram (8,16), the top is paramagnetic while the bottom is (just) in the ordered phase, consistent with the neutron scattering findings.

The resistance, specific heat and susceptibility measurements were performed on two pieces of the sample of sizes 1.2mm x 0.85mm x 6.0 mm, weighing 50 mg, taken from the top part of the sample. The resistance data were taken using an Oxford Instruments MagLab for the temperature range 2K- 300K (shown in black in Fig. S3), while an Oxford Instruments Heliox VL was used for the range 0.3K- 15K (shown in blue).

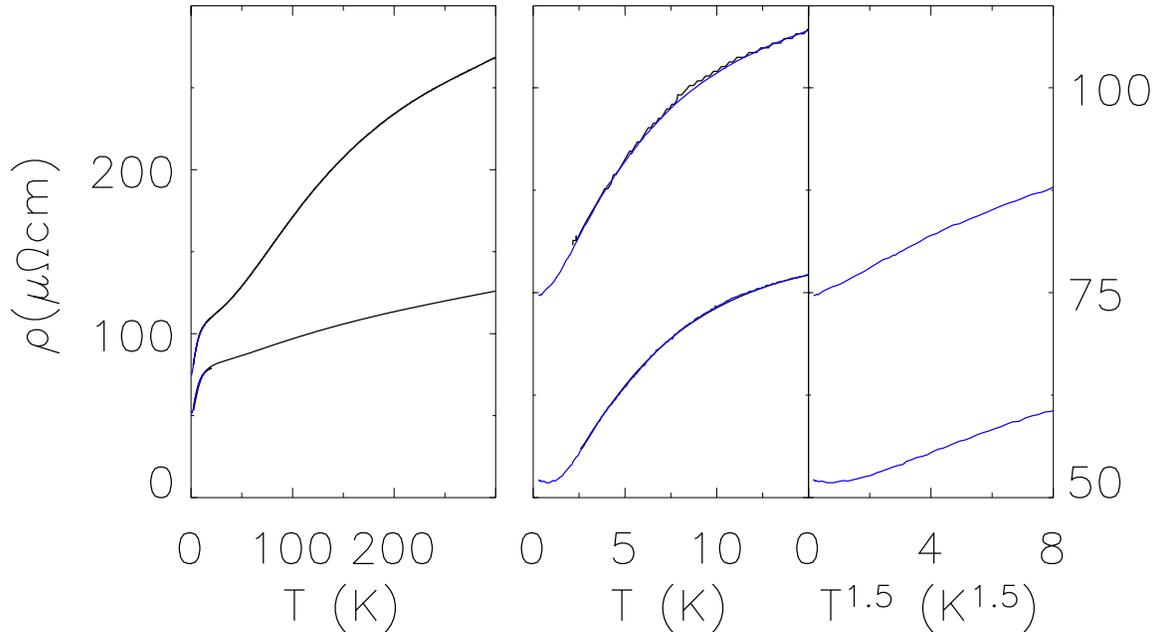

Fig. S3: Resistivity measured along the [100] direction (top curves) and along the [001]-direction (bottom curves). The black curves have been measured using an Oxford Instruments Maglab, the blue curves using the Heliox VL. The data measured on the two systems show good overlap.

The electrical contacts were placed in the standard 4-probe geometry, using Cu wires and silver paste. The ac-currents were along [100] (upper curves in fig. S3, $\rho_0$ = 74.3 $\mu\Omega$cm) and [001] (lower curves, $\rho_0 \approx$ 52 $\mu\Omega$cm). While the room temperature resistivity values are similar to those measured (S1) in pure $CeFe_2Ge_2$ (250 and 85 $\mu\Omega$cm along [100] and [001], respectively), the $\rho_0$–values are considerably larger, reflecting disorder due to Ru substitution on Fe-sites.

The data show the onset of coherence (1) around T~ 15 K. The resistivity data taken with the current along [001] bottom out at T= 0.8 K, below which they show a small increase. The data along [100] do not show this increase. The [100]-data vary with temperature as $T^{1.5}$ in the range 1.5 K < T < 3.5 K, with a higher exponent (1.86) below 1.5 K, and a lower exponent above 3.5 K.

We measured the specific heat on one of the 50 mg pieces cut from the top of the crystal. The data were collected using Quantum Design Physical Properties Measurement Systems, the magnetic contribution was isolated by subtracting the specific heat values for a non-magnetic isostructural polycrystalline reference sample ($LaFe_2Ge_2$). The data show non-Fermi-liquid behavior (1) over a large temperature range, but similar to the results for the resistivity, we find at the lowest temperatures that the data show signs of the sample being slightly on the paramagnetic side of the QCP. The coefficient of the linear term in the specific heat at T= 0.3 K is $\gamma = (748 \pm 5)$ mJ/mol.K$^2$.

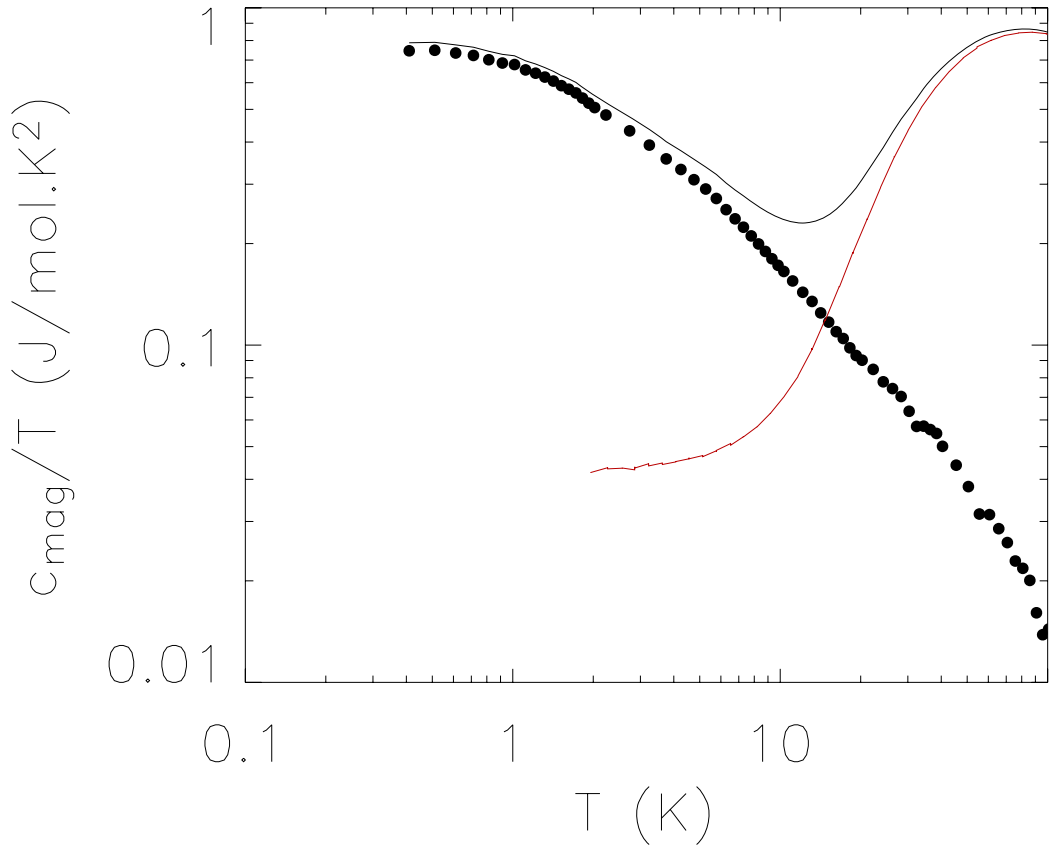

Fig. S4: Magnetic contribution to the $Ce(Ru_{1-x}Fe_x)_2Ge_2$ specific heat. The solid black line are the data measured for the 50 mg single crystal, the solid red line are the data measured for a 9 mg polycrystalline sample of $LaFe_2Ge2$. The black dots are the difference between the two data sets (the reference data have been extrapolated for T < 2 K)

The susceptibility was determined using a Quantum Design SQUID magnetometer in the temperature range 3.5 K-300 K. We used two 50 mg samples, taken from the top part of the crystal. We collected data in a field of 0.2 T (the magnetization was found to be linear in fields up to 0.5 T). The data shown in Fig. S5 confirm that the [001]-axis is the easy axis.

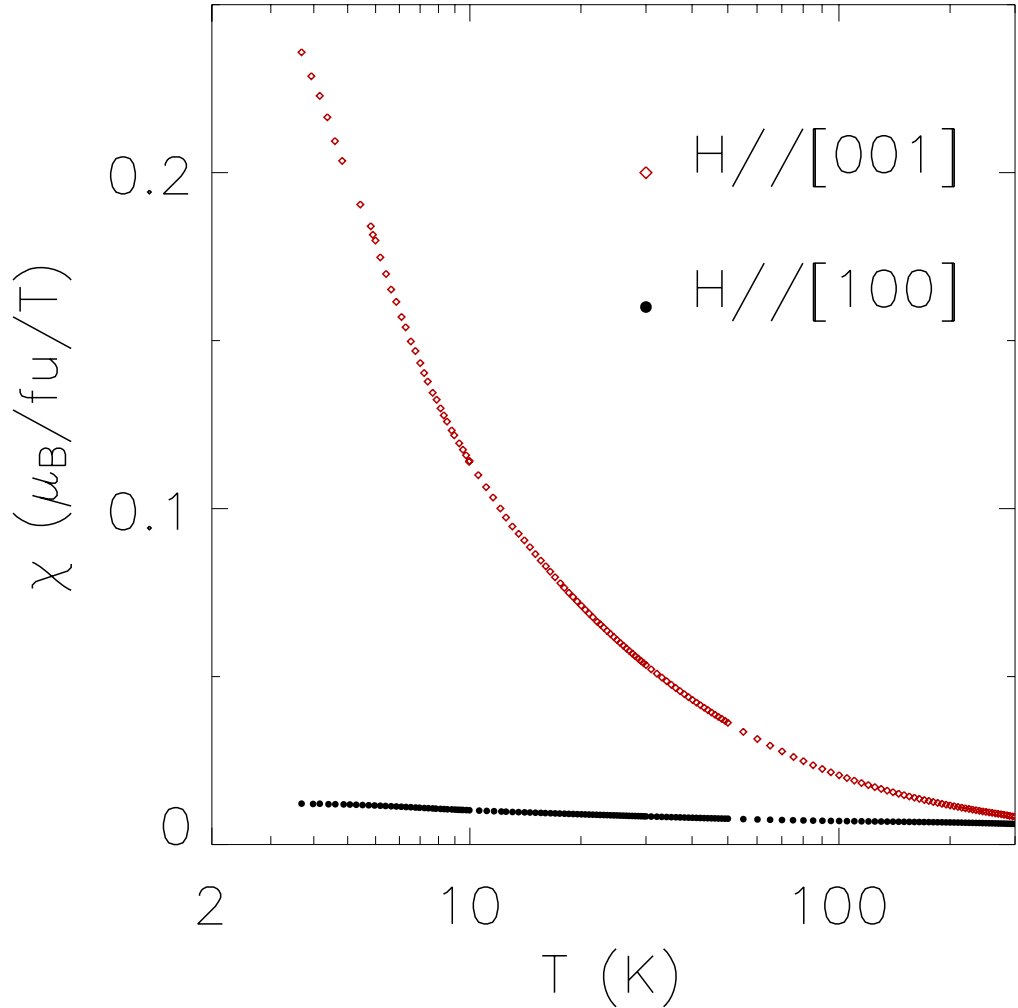

Fig. S5: Ce(Ru$_{1-x}$Fe$_x$)$_2$Ge$_2$ susceptibility measured in a magnetic field of 0.2 T.

Neutron scattering: data reduction and size and direction of the ordered moment

Graphite filters (3") were used for the HB3 experiment to completely eliminate higher order scattering, the DCS experiment did not require any filters. Empty cryostat runs were performed for both experiments, and the contribution of the cryostat to the scattering was found to be very small. We did not attempt to correct for the angular dependence of the self-attenuation [self-shielding by the sample because of scattering and absorption of neutrons] because the attenuation is expected to be virtually angle-independent because of the cylindrical geometry of the sample. Normalization to a nuclear Bragg peak, or to the incoherent scattering by the sample, both of which are subject to the same self-attenuation factor, effectively corrects for the average level of self-attenuation. Based on the temperature dependence of polycrystalline data taken on an isostructural non-magnetic sample (19), we conclude, for temperatures T< 56 K and for wave vectors near (1,1,0.45), that multiphonon effects are absent. This is corroborated by multiphonon calculations using the measured phonon dispersion near (1,1,0).

In this paper we have used the signal at T= 56 K as our background signal. The 56 K signal contains the nuclear scattering, the almost negligible scattering by the cryostat, and also the magnetic scattering associated with rapidly fluctuating, uncorrelated cerium moments. The first two contributions are almost temperature independent below T= 56 K, and the latter contribution does not show any angular dependence at 56 K. However, when we subtract this isotropic magnetic contribution it gives rise to a directional dependent background, as can be seen in Fig. 2. The reason for this is straightforward: first, when the temperature is lowered, the conduction electrons do a better job of shielding the Ce moments reducing the overall amount of magnetic scattering. Second, when the magnetic intensity starts to peak at (n,n,2m±0.45), intensity is removed from other parts of reciprocal space, most notable from (n/2,n/2,2m±0.45) where the ordered moments add up in anti-phase. This leads to a negative signal after subtracting the 56 K data from the 2 K data for regions near (n/2,n/2,2m±0.45).

Within the limitations stated above, using the T= 56 K scattered signal as the background signal to remove incoherent scattering and the neutrons scattered by the Al sample holder works very well at (1,1,0.45), but it runs into some difficulties for higher momentum transfers: the crystal very slightly contracts between T= 2 K and T= 56 K, with the result that the nuclear Bragg peaks undergo a small shift in scattering angle. While this shift is small, it has a noticeable effect when subtracting two Bragg peaks. In order to arrive at the data in Fig. 1, we included a 0.15% contraction for this temperature range (based on the shifted Bragg positions). In view of the decreasing magnetic signal as a function of increased momentum transfer, the uncertainties inherent to this correction become more pronounced, and we could not reliably determine the magnetic signal beyond q=(1,1,6).

The magnetic neutron scattering cross section depends on the angle between the cerium moment direction and the direction of momentum transfer, allowing us to determine that the moments are aligned along the c-axis and that they are partially shielded by the conduction electrons. This is in agreement with the uniform susceptibility data presented in Fig. S5. Consistent with our findings that the moments order along the c-axis, neutron scattering data (8,19) on polycrystalline quantum critical $Ce(Ru_{1-x}Fe_x)_2Ge_2$ showed the appearance of satellite peaks around (1,0,1), but not along (0,0,2). From our present neutron scattering experiments we calculate the size of the ordered moment to be ~0.12-0.2 $\mu_B$ per cerium ion at T= 2 K. The uncertainty stems from the fact that the increased scattered intensity associated with the onset of ordering is spread out around the ordering wave vector with the result that some of this intensity is not intercepted by the detector (which has a finite height). We can estimate the size of the ordered moment by comparing the intensity of the (1,1,0) nuclear Bragg peak [which is so sharp that the detector intercepts all the intensity] to the integrated intensity of the magnetic scattering, correcting for finite size effects as best as possible. This yields an ordered moment of (0.15±0.03) $\mu_B$ per cerium ion. Alternatively, we can use the incoherent scattering for the normalization. Again, this leads to similar problems since the incoherent scattering does not show any angular dependence, whereas the magnetic scattering does. The incoherent method yields (0.18±0.03) $\mu_B$ per cerium ion.

The DCS data and data taken on the polycrystalline samples (8) are in good agreement. Using the DCS data, we can integrate the total scattering (after normalization to the incoherent scattering) between –10 meV < E < 2 meV to find the strength of the magnetic scattering away from ordering (the q-independent contribution associated with the local susceptibility). We find the scattering power to be 0.5±0.05 $\mu_B^2$ per formula unit at 56 K, and 0.3±0.05 $\mu_B^2$ per formula unit at 0.4 K. This is in good agreement with the results reported (8) for the polycrystalline material. The HB3 experiments show that there is considerable scattering power at higher energy transfers (E> 2meV).

Magnetic form factor

Polycrystalline data (19) in the region $q < 2\text{Å}^{-1}$ had shown that the magnetic intensity, corresponding to the local moments at T= 200 K, did not show any significant q-dependence in this q-range. Had the moments been located on the transition metal ion, a decline in magnetic scattering intensity by about 40% would have been observed in this q-range. Thus, at high temperatures the moments are known to reside of the cerium-ions, however, this does not automatically imply that it is the selfsame cerium moments that order. Fig. 1 demonstrates that the ordered moments are indeed also located on the cerium ions. The f-electron form factor was estimated from the detailed measurements (21) on the cerium form factor in $CeRh_3B_2$, the d-electron form factor was calculated by Fourier transformation of the radial dependence of the Fe d-orbitals using an effective nuclear charge $Z_{eff}$=12. The detailed shape of the form factor (which requires knowledge of the crystal electric fields) is not important in reaching the conclusion that the magnetic scattering originates from f-electrons.

Lineshapes

We fitted our data to a Lorentzian line shape (excluding the two data points where an ordered component appeared, see Fig. 2) convoluted with the resolution function determined from grid scans around the (1,1,0) Bragg peak. At T= 2 K, the quality of the fits is very good: $\chi^2 \leq 1$. For all temperatures, the FWHM of the resolution function was less than 20% of the magnetic signal width rendering the determination of the correlation lengths unambiguous. Leaving the resolution width as an adjustable parameter, representing a mixture of various concentrations, did not improve the quality of the fits. Other lineshapes, such as fitting to the Fisher-Burford form (S2), also did not improve the quality of the fit at T= 2 K (yielding $\eta$= 0.08 ± 0.16). From this we conclude that a Lorentzian line shape best represents the decay of the fluctuations, and that the narrow range of concentrations *x* present in our sample as determined by electron microprobe do not alter the observed lineshape. At the higher temperatures, the line widths are so large that a sizeable portion of the magnetic intensity interferes with the nuclear Bragg peak, and Al powder peaks. Combined with the uncertainty in background due to the sample contraction, we can no longer determine very accurate linewidths from the data at the higher temperatures.

Zero-point motion amplitude

We performed neutron powder diffraction experiments at the Missouri Research Reactor on a pulverized piece of quantum critical $Ce(Ru_{1-x}Fe_x)_2Ge_2$ [the same sample that had previously been studied (8)]. The spectrometer was operated using an incident neutron wavelength $\lambda= 1.48$ Å ($2\theta_{max}= 108°$), and powder patterns were collected at T= 16, 50, 100 and 300 K. Standard Rietveld refinements were performed to obtain the cerium anisotropic Debye-Waller factors along the a and c-axes [$<u^2_a> = <u^2_b> \neq <u^2_c>$]. The results show that the cerium-displacements $\sqrt{<u^2>}$ are indeed of the order of 0.1 Å. We find $<u^2_c> = (7\pm1)\times10^{-3}$ Å$^2$ for all 4 temperatures, and $<u^2_a>= (12\pm2)$, $(7\pm2)$, $(3\pm1)$, $(10\pm2)$ $\times10^{-3}$ Å$^2$ for T= 300, 100, 50 and 16 K, respectively. We do not understand why $<u^2_a>$ shows an increase at 16 K; we will perform measurements spanning a larger range of momentum transfers and temperatures, in here we just use the results to illustrate the importance of zero-point-motion even at the lowest temperatures.

Zero-point-motion versus concentrational fluctuations: which one is more important?

Long-range order is expected when there is a path linking moments on either side of the sample. First, this rules out the role of sample defects: it seems inconceivable that vacancies or impurities in the sample could conspire in such a way so as to make such a path impossible.
Both types of fluctuations would in principle account for the observed symmetry of the correlations in $Ce(Ru_{1-x}Fe_x)_2Ge_2$. $CeRu_2Ge_2$ has lattice constants (16) of a=4.26 Å and c=10.0 Å, those for $CeFe_2Ge_2$ are a=4.08 Å and 10.5 Å. Thus, both substitution and zero-point-motion change distances on length scales of the order of 0.1 Å, sufficient to significantly alter the overlap between localized and extended electron states. Also, a 1 in 4 substitution of Fe atoms by Ru atoms could well make the first percolation path possible (assuming that Ru substitution allows nearby cerium ions to develop a permanent moment). Thus, both types of fluctuations could be the root cause of the inferred local, random variation of the coupling between the local moments and the conduction electrons. Note that substitution effects by themselves cannot be the explanation for the observed E/T-scaling (7) in $CeAu_{5.9}Cu_{0.1}$ as the level of substitution is not sufficient to upset the percolation path. However, there is some evidence (7,9) in $CeAu_{5.9}Cu_{0.1}$ that the spin fluctuations have a reduced dimensionality, therefore the percolation arguments presented for our sample might not necessarily hold for $CeAu_{5.9}Cu_{0.1}$. It would be interesting to see what the correlation symmetry is in $CeNi_2Ge_2$, a stoichiometric sample (2) that is at a comparable position in the phase diagram as the sample we have investigated. In stoichiometric $CeNi_2Ge_2$, only zero-point motion should drive the system away from the ordered phase. Should this be accompanied by E/T-scaling, then we can conclude that E/T-scaling originates from zero-point motion; conversely, should E/T-scaling not be observed, then our results identify substitution effects (local disorder) as the ultimate origin of E/T-scaling.

In all, the sample symmetry and composition turned out to be just right for the experiments. The fact that the crystal symmetry was non-cubic allowed us to notice that the symmetry of the fluctuations was cubic. The small part of the sample that is Fe-deficient confirmed that we are very close to ordering without unduly affecting the interpretation of the data, while being slightly paramagnetic ensured that we could follow the magnetic signal down to the temperature at which quantum fluctuations dominate. The data also confirmed that the magnetic moments do indeed survive down to the lowest temperatures, even in the paramagnetic heavy-fermion phase. Hence, Kondo screening is never complete, something that we ascribe to the zero-point motion of the cerium ions.